\documentclass{amsart}
\usepackage{amssymb, latexsym}
\begin{document}
\title{\bf Coulomb Potential of a Point Mass in Theta Noncommutative Geometry}
\author{A. Lewis Licht}
\address{Dept. of Physics\\ University of Illinois at Chicago\\ 
Chicago, Illinois 60607\\ licht@uic.edu}

\begin{abstract}
    We investigate the form of the Coulomb potential of a point charge in a
    noncommutative geometry, using a state of minimal dispersion. We 
    find the deviation of the potential at large distances 
    from the point, distinguishing between coordinate distance and measured 
    distance. Defining the ``effective'' value of an operator as its 
    expectation value in a minimum dispersion state centered at a 
    point, we find the effective potential to be finite at the 
    origin, the effective charge density to be Gaussian and the 
    effective total electrostatic energy to be finite. However, the true 
    total electrostatic energy operator is shown to still be infinite.
\end{abstract}
\maketitle
\section{Introduction}\label{S:intro}
We consider the form of noncommutative geometry where the operators 
that measure position, the ``physical'' position operators, satisfy
\begin{equation}\label{E:noncomtheta}
    \left[ {x^i ,\;x^j } \right] = \theta ^{ij} 
\end{equation}

Where $\theta^{i,j}$ is a constant antisymmetric tensor, and where 
the indexes i, j = 1, 2, 3. The components $\theta^{0,i}$ are assumed 
to vanish. The physical momentum operators are assumed to have the 
conventional commutation relations~\cite{douglas}
\begin{equation}\label{E:comconven}
 \begin{array}{c}
 \left[ {p^i ,\;p^j } \right] = 0 \\ 
 \left[ {x^i ,\;p^j } \right] = i\delta ^{ij}  \\ 
 \end{array}
 \end{equation}
 
We investigate for such a geometry the problem of measuring distance 
from a point.  We also investigate the operational definition of the 
value at a point of such quantities as the Coulomb potential, the 
charge distribution and the electric field.

Let $\left| {\Psi _0 } \right\rangle $
denote a point-like state at the origin.  This could be a wave packet 
spherically symmetric about the origin in commutative theory, or as 
spherically symmetric a state as possible in NC theory.  The 
spherical symmetry implies
\begin{equation}\label{E:symm}
\left\langle {\Psi _0 \left| {{\bf{x}}\left| {\Psi _0 } \right.} \right.} \right\rangle  = 0
\end{equation}
We translate the state $\left| {\Psi _0 } \right\rangle $ to a point a distance 
${\bf{a}}$ from the origin, obtaining the state
\begin{equation}
\left| {\Psi _0 \left( {\mathbf{a}} \right)} \right\rangle  = \exp \left( { - i{\mathbf{a}} \cdot {\mathbf{p}}} \right)\left| {\Psi _0 } \right\rangle 
\end{equation}

The vector ${\bf{a}}$ is a coordinate displacement.  Its length $a$ is 
not actually the displacement distance that would be measured.  The 
measured distance is
\begin{equation}
\begin{gathered}
  \left\langle r \right\rangle  = \left\langle {\Psi _0 \left( {\mathbf{a}} \right)} \right|\sqrt {{\mathbf{x}} \cdot {\mathbf{x}}} \left| {\left| {\Psi _0 \left( {\mathbf{a}} \right)} \right.} \right\rangle  \\ 
   = \left\langle {\Psi _0 } \right|\sqrt {{\mathbf{x}} \cdot {\mathbf{x}} + 2{\mathbf{x}} \cdot {\mathbf{a}} + a^2 } \left| {\Psi _0 } \right\rangle  \\ 
   = a + \left\langle {\Psi _0 \left| {{\mathbf{x}} \cdot {\mathbf{\hat n}} + \frac{1}
{{2a}}\left( {{\mathbf{x}}^2  - \left( {{\mathbf{x}} \cdot {\mathbf{\hat n}}} \right)^2 } \right)} \right|\Psi _0 } \right\rangle  + O\left( {\frac{1}
{{a^2 }}} \right) \\ 
\end{gathered} 
\end{equation}
    
Where ${\mathbf{\hat n}}$ is the unit vector along ${\mathbf{a}}$.  
Using Eq.~(\ref{E:symm}), this becomes
\begin{equation}\label{E:dist}
\left\langle r \right\rangle  = a + \frac{1}
{{2a}}\left\langle {\Psi _0 \left| {\left( {{\mathbf{x}}^2  - \left( {{\mathbf{x}} \cdot {\mathbf{\hat n}}} \right)^2 } \right)} \right|\Psi _0 } \right\rangle  + O\left( {\frac{1}
{{a^2 }}} \right)
\end{equation}
The Coulomb potential that would be measured at the origin due to a 
point particle at the coordinate position ${\bf{a}}$ is 

\begin{equation}\label{E:coulpota}
\begin{gathered}
  \left\langle {\frac{1}
{r}} \right\rangle  = \left\langle {\Psi _0 \left( {\mathbf{a}} \right)} \right|\frac{1}
{{\sqrt {{\mathbf{x}}^2 } }}\left| {\Psi _0 \left( {\mathbf{a}} \right)} \right\rangle  \\ 
   = \left\langle {\Psi _0 } \right|\frac{1}
{{\sqrt {{\mathbf{x}}^2  + 2{\mathbf{x}} \cdot {\mathbf{a}} + a^2 } }}\left| {\Psi _0 } \right\rangle  \\ 
   = \frac{1}
{a} - \left\langle {\Psi _0 \left| {\frac{1}
{{2a^3 }}\left( {{\mathbf{x}}^2  - 3\left( {{\mathbf{x}} \cdot {\mathbf{\hat n}}} \right)^2 } \right)} \right|\Psi _0 } \right\rangle  + O\left( {\frac{1}
{{a^4 }}} \right) \\ 
\end{gathered} 
\end{equation}

This is however expressed in terms of coordinate distance 
${\bf{a}}$.  Eq.~(\ref{E:dist}) may be used to express it in terms of 
the observed distance $\left\langle r \right\rangle $
\begin{equation}\label{E:obsvcoupot}
\left\langle {\frac{1}
{r}} \right\rangle  =  = \frac{1}
{{\left\langle r \right\rangle }} + \frac{1}
{{\left\langle r \right\rangle ^3 }}\left\langle {\Psi _0 \left| {\left( {{\mathbf{x}} \cdot {\mathbf{\hat n}}} \right)^2 } \right|\Psi _0 } \right\rangle  + O\left( {\frac{1}
{{\left\langle r \right\rangle ^4 }}} \right)
\end{equation}

For any operator A, the expectation value in the translated state
\begin{equation}
\begin{gathered}
  \left\langle {\Psi _0 \left( {\mathbf{a}} \right)} \right|A\left| {\Psi _0 \left( {\mathbf{a}} \right)} \right\rangle  = \left\langle {\Psi _0 } \right|\exp \left( { + i{\mathbf{a}} \cdot {\mathbf{p}}} \right)A\exp \left( { - i{\mathbf{a}} \cdot {\mathbf{p}}} \right)\left| {\Psi _0 } \right\rangle  \\ 
   = \left\langle {\Psi _0 } \right|A\left( {\mathbf{a}} \right)\left| {\Psi _0 } \right\rangle  \\ 
\end{gathered} 
\end{equation}
can also be interpreted as measuring the operator A at the 
``coordinate point'' ${\bf{a}}$ in the compact state $\Psi _0$.  We 
interprete this as the ``effective'' value of the operator A at 
${\bf{a}}$ in the state $\Psi _0$.

We examine measured distance and Coulomb potential 
for a commutative geometry in Section (2) and for $\theta$ NC 
geometry in Section (3).  In Section (4) it is shown that in $\theta$ NC 
geometry the effective Coulomb potential, as seen from a compact 
state, is finite at the origin.  The effective charge density is 
derived in Section (5), and the effective elective field in Section 
(6).  The total ``effective'' electrostatic energy is calculated and 
found to be finite.  However, in Section (7) we consider the true 
electric field operator and show that the integral over all space of 
the corresponding electrostatic energy density can be defined in $\theta$ NC 
geometry and is infinite.

\section{Commutative Geometry}\label{S:comm}
We consider here an ordinary free particle in a Gaussian wasve packet 
of coordinate dispersion $\sigma $.  The wave function in x space is
\begin{equation}
\Psi \left( {\mathbf{x}} \right) = \frac{1}
{{\left( {2\pi \sigma } \right)^{\tfrac{3}
{4}} }}\exp \left( { - \frac{{{\mathbf{x}}^2 }}
{{4\sigma }}} \right)
\end{equation}

For this particle,
\begin{equation}
\begin{gathered}
  \left\langle {{\mathbf{x}}^2 } \right\rangle  = 3\sigma  \\ 
  \left\langle {\left( {{\mathbf{x}} \cdot {\mathbf{\hat n}}} \right)^2 } \right\rangle  = \sigma  \\ 
\end{gathered} 
\end{equation}
Then the measured distance is
\begin{equation}
\left\langle r \right\rangle  = a + \frac{\sigma }
{a} + O\left( {\frac{1}
{{a^2 }}} \right)
\end{equation}
and the Coulomb potential is
\begin{equation}
\left\langle {\frac{1}
{r}} \right\rangle  = \frac{1}
{{\left\langle r \right\rangle }} + \frac{\sigma }
{{\left\langle r \right\rangle ^3 }} + O\left( {\frac{1}
{{\left\langle r \right\rangle ^4 }}} \right)
\end{equation}

Of course, for this system $\sigma$ can be made arbitrarily small, 
and the Coulomb potential arbitrarily close to 1/r .

\section{Theta Noncommutativity}\label{S:thetnc}

The operators
\begin{equation}
\xi ^i  = x^i  + \frac{{\theta ^{ik} }}
{2}p^k 
\end{equation}
have the nice property that
\begin{equation}
\begin{gathered}
  \left[ {\xi ^i ,\;\xi ^j } \right] = 0 \\ 
  \left[ {\xi ^i ,\;p^j } \right] = i\delta ^{ij}  \\ 
\end{gathered} 
\end{equation}
The $\xi ^i $ can be considered as hypothetical coordinate 
operators, where the $x^i $ are the operators that correspond to 
actual physical measurements of position.

The anti-symmetry of the $\theta ^{ij} $matrix allows us to introduce 
a vector ${\mathbf{\theta}}$ such that
\begin{equation}
\begin{gathered}
  \theta ^{ij}  = \varepsilon _{ijk} \theta ^k  \\ 
  \theta ^k  = \frac{1}
{2}\varepsilon _{ijk} \theta ^{ij}  \\ 
\end{gathered} 
\end{equation}
Then

\begin{equation}
{\mathbf{x}} = {\mathbf{\xi}}  + \tfrac{1}
{2}{\mathbf{\theta}}  \times {\mathbf{p}}
\end{equation}
and the distance squared operator becomes

\begin{equation}
{\mathbf{x}}^2  = {\mathbf{\xi}} ^2  - {\mathbf{\theta}}  \cdot 
{\mathbf{\xi}}  \times {\mathbf{p}} + \frac{{{\mathbf{\theta}} ^2 }}
{4}\left( {{\mathbf{p}}^2  - \left( {{\mathbf{p}} \cdot {\mathbf{\theta}} } \right)^2 } \right)
\end{equation}

We find this operator's spectrum by introducing creation and 
annihilation operators:

\begin{equation}
\begin{gathered}
  u_\alpha   = \frac{1}
{{\sqrt \theta  }}\xi _\alpha   + i\frac{{\sqrt \theta  }}
{2}p_\alpha   \\ 
  \zeta  = \xi _3  \\ 
\end{gathered} 
\end{equation}

where $\alpha = 1,2$.  Then

\begin{equation}\label{E:xsquar}
{\mathbf{x}}^2  = \zeta ^2  + i\theta \left( {u_1^\dag  u_2  - u_2^\dag  u_1 } \right) + \theta \left( {u_1^\dag  u_1  + u_2^\dag  u_2  + 1} \right)
\end{equation}

With the substitution

\begin{equation}
\left( {\begin{array}{*{20}c}
   {u_1 }  \\
   {u_2 }  \\
 \end{array} } \right) = \frac{1}
{{\sqrt 2 }}\left( {\begin{array}{*{20}c}
   1 & 1  \\
   i & { - i}  \\
\end{array} } \right)\left( {\begin{array}{*{20}c}
   a  \\
   b  \\
\end{array} } \right)
\end{equation}

Where a and b are annihilation operators, Eq.~(\ref{E:xsquar}) becomes

\begin{equation}
{\mathbf{x}}^2  = \zeta ^2  + \theta \left( {2b^\dag  b + 1} \right)
\end{equation}
The angular momentum operator ${\mathbf{\theta}}  \cdot 
{\mathbf{\xi}}  \times {\mathbf{p}}$ corresponding to rotations about 
the 3-axis is now
\begin{equation}
{\mathbf{\theta}}  \cdot 
{\mathbf{\xi}}  \times {\mathbf{p}} = a^\dag  a - b^\dag  b
\end{equation}

The general basis state can be written as 

\begin{equation}
\left| {\Psi _{nm} \left( \zeta  \right)} \right\rangle  = \psi \left( \zeta  \right)\left| {nm} \right\rangle 
\end{equation}

where $\psi \left( \zeta  \right)$ is an arbitrary wavefunction, and 
$\left| {nm} \right\rangle$ is a harmonic oscillator basis state, with

\begin{equation}
\begin{gathered}
  \left| {nm} \right\rangle  = \frac{{a^{\dag n} b^{\dag m} }}
{{\sqrt {n!m!} }}\left| {00} \right\rangle  \\ 
  a\left| {00} \right\rangle  = 0 \\ 
  b\left| {00} \right\rangle  = 0 \\ 
\end{gathered} 
\end{equation}

To get the most compact state, with maximum symmetry, so that 
Eq.~(\ref{E:symm})is satisfied, we take $\psi \left( \zeta  \right)$ 
to be a gaussian with dispersion $\sigma$, and n = m = 0.  Then our 
standard state has the wavefunction

\begin{equation}
\left\langle {{\mathbf{\xi}} _ \bot  ,\;\zeta \left| {\Psi _{00} } \right.} \right\rangle  = \frac{1}
{{\left( {2\pi } \right)^{\frac{3}
{4}} \sigma ^{\frac{1}
{4}} \left( {\frac{\theta }
{4}} \right)^{\frac{1}
{2}} }}\exp \left[ { - \frac{{{\mathbf{\xi}} _ \bot  ^2 }}
{\theta } - \frac{{\zeta ^2 }}
{{4\sigma }}} \right]
\end{equation}

Then with ${\mathbf{\hat n}} = \left( {\alpha ,\;\phi } \right)$, 
we get

\begin{equation}
\begin{gathered}
  \left\langle {{\mathbf{x}}^2 } \right\rangle  = \sigma  + \theta  \\ 
  \left\langle {\left( {{\mathbf{x}} \cdot {\mathbf{\hat n}}} \right)^2 } \right\rangle  = \sigma \cos ^2 \alpha  + \frac{\theta }
{2}\sin ^2 \alpha  \\ 
\end{gathered} 
\end{equation}

and

\begin{equation}
\left\langle {\frac{1}
{r}} \right\rangle  = \frac{1}
{{\left\langle r \right\rangle }} + \frac{1}
{{\left\langle r \right\rangle ^3 }}\left( {\sigma \cos ^2 \alpha  + \frac{\theta }
{2}\sin ^2 \alpha } \right) + O\left( {\frac{1}
{{\left\langle r \right\rangle ^4 }}} \right)
\end{equation}

For this system, the deviation from the 1/r Coulomb potential depends 
on the direction along which the potential is measured. In terms of 
the coordinate distance a, this is

\begin{equation}\label{E:acoul}
\left\langle {\frac{1}
{r}} \right\rangle  = \frac{1}
{a} + \frac{1}
{{2a^3 }}\left( {\sigma \left( {3\cos ^2 \alpha  - 1} \right) + \frac{\theta }
{2}\left( {3\sin ^2 \alpha  - 2} \right)} \right) + O\left( {\frac{1}
{{a^4 }}} \right)
\end{equation}

\section{Effective Potential at Origin}\label{S:coordpos}

We distinguish the vector operator that measures position, 
${\mathbf{x}}$, from the auxiliary vector operator ${\mathbf{\xi}}  = 
\left( {{\mathbf{\xi}} _ \bot  ,{\kern 1pt} \,\zeta } \right)$.  
Eigenvalues of the latter we will consider as a kind of 
``coordinate'' position, analogous to the coordinate position of 
general relativity.  Expectation values of the operator ${\mathbf{x}}$ 
give the ``physical'' position.

It is of interest ot ask what an operator, in particular the Coulomb 
potential, looks like in terms of coordinate position.  A qualitative treatment of this problem 
has been given by Colatto et al.~\cite{colatto}  To determine 
this, we consider the states

\begin{equation}
\left| {\Psi \left( {\sigma ,\;\sigma _ \bot  ,\;{\mathbf{a}}} 
\right)} \right\rangle  = \int {d^3 \xi } \left|{\mathbf{ \xi}}  \right\rangle \frac{1}
{{\left( {2\pi } \right)^{\frac{3}
{4}} \sigma ^{\frac{1}
{4}} \sigma _ \bot ^{\frac{1}
{2}} }}\exp \left[ { - \frac{{\left( {\zeta  - a^3 } \right)^2 }}
{{4\sigma }} - \frac{{\left( {{\mathbf{\xi}} _ \bot   - {\mathbf{a}}_ \bot  } \right)^2 }}
{{4\sigma _ \bot  }}} \right]
\end{equation}

These states give us a kind of window at the point ${\mathbf{a}}$, of 
widths $\sqrt \sigma  ,\;\sqrt {\sigma _ \bot  } $ in the z, r 
directions, to determine the ``effective'' value of any operator A at that point:

\begin{equation}
\begin{gathered}
  \left\langle {A\left( {\mathbf{a}} \right)} \right\rangle  = \left\langle {\Psi \left( {\sigma ,\;\sigma _ \bot  ,\;{\mathbf{a}}} \right)} \right|A\left| {\Psi \left( {\sigma \;,\sigma _ \bot  ,\;{\mathbf{a}}} \right)} \right\rangle  \\ 
   = \left\langle {\mathbf{a}} \right|A\left| {\mathbf{a}} \right\rangle  \\ 
\end{gathered} 
\end{equation}

To look at the Coulomb potential we use the Fourier transform 
expression

\begin{equation}
\frac{1}
{r} = \frac{1}
{{2\pi ^2 }}\int {d^3 q} \frac{1}
{{q^2 }}\exp \left[ {i{\mathbf{q}} \cdot {\mathbf{x}}} \right]
\end{equation}

We note that

\begin{equation}
\begin{gathered}
  \exp \left[ {i{\mathbf{q}} \cdot {\mathbf{x}}} \right] = \exp 
  \left[ {i{\mathbf{q}} \cdot \left( {{\mathbf{\xi}}  + \tfrac{1}
{2}{\mathbf{\theta}}  \times {\mathbf{p}}} \right)} \right] \\ 
   = \exp \left[ {i{\mathbf{q}} \cdot {\mathbf{\xi}} } \right]\exp \left[ {i{\mathbf{q}} \cdot \tfrac{1}
{2}{\mathbf{\theta}}  \times {\mathbf{p}}} \right] \\ 
\end{gathered} 
\end{equation}

since

\begin{equation}
\left[ {{\mathbf{q}} \cdot {\mathbf{\xi}} ,\;{\mathbf{q}} \cdot 
{\mathbf{\theta}}  \times {\mathbf{p}}} \right] = i{\mathbf{q}} \cdot {\mathbf{\theta}}  \times {\mathbf{q}} = 0
\end{equation}

also

\begin{equation}
\exp \left[ {i{\mathbf{q}} \cdot \tfrac{1}
{2}{\mathbf{\theta}}  \times {\mathbf{p}}} \right]\left| {\mathbf{\xi}}  \right\rangle  = \left| {{\mathbf{\xi}}  + \tfrac{1}
{2}{\mathbf{\theta}}  \times {\mathbf{q}}} \right\rangle 
\end{equation}

Then

\begin{equation}
\begin{aligned}
  \left\langle {\mathbf{a}} \right|\frac{1}
{r}\left| {\mathbf{a}} \right\rangle  =  & \frac{1}
{{2\pi ^2 \left( {2\pi } \right)^{\frac{3}
{2}} }}\int {\frac{{d^3 \xi 'd^3 \xi }}
{{\sigma _ \bot  \sqrt \sigma  }}\frac{{d^3 q}}
{{q^2 }}\exp \left[ {i{\mathbf{q}} \cdot {\mathbf{\xi}} ' - \frac{{\left( {\zeta ' - a^3 } \right)^2  + \left( {\zeta  - a^3 } \right)^2 }}
{{4\sigma }} - \frac{{\left( {{\mathbf{\xi}} '_ \bot   - {\mathbf{a}}_ \bot  } \right)^2  + \left( {{\mathbf{\xi}} _ \bot   - {\mathbf{a}}_ \bot  } \right)^2 }}
{{4\sigma _ \bot  }}} \right]}  \\ 
   &  \times \left\langle {\mathbf{\xi}} '\vert \, {\mathbf{\xi}}  + \frac{1}
{2}{\mathbf{\theta}}  \times {\mathbf{q}}  \right\rangle  \\ 
\end{aligned} 
\end{equation}

which becomes

\begin{equation}\label{E:potina}
\left\langle {\mathbf{a}} \right|\frac{1}
{r}\left| {\mathbf{a}} \right\rangle  = \frac{1}
{{2\pi ^2 }}\int {\frac{{d^3 q}}
{{q^2 }}\exp \left[ {i{\mathbf{q}} \cdot {\mathbf{a}} - \frac{{\sigma \left( {q^3 } \right)^2 }}
{2} - \frac{{\sigma _p {\mathbf{q}}_ \bot ^2 }}
{2}} \right]} 
\end{equation}

where

\begin{equation}\label{E:sigmap}
\sigma _p  = \sigma _ \bot   + \frac{{\theta ^2 }}
{{16\sigma _ \bot  }}
\end{equation}

At very large $\left| a \right|$, we can write this to first order in the 
dispersions,

\begin{equation}\label{E:largea}
\begin{gathered}
  \left\langle {\mathbf{a}} \right|\frac{1}
{r}\left| {\mathbf{a}} \right\rangle  = \left[ {1 + \frac{\sigma }
{2}\left( {\frac{\partial }
{{\partial a^3 }}} \right)^2  + \frac{{\sigma _p }}
{2}\nabla _ \bot ^2 } \right]\frac{1}
{a} \\ 
   = \frac{1}
{a} + \frac{1}
{{a^3 }}\left[ {\sigma \left( {3\cos ^2 \alpha  - 1} \right) + \sigma _p \left( {3\sin ^2 \alpha  - 2} \right)} \right] \\ 
\end{gathered} 
\end{equation}

The dispersion of Eq.~(\ref{E:sigmap}) is minimized by $\sigma _ \bot   = \frac{\theta }
{4}$ , which makes $\sigma _p  = \frac{\theta }
{2}$ and Eq.~(\ref{E:largea}) becomes identical with 
Eq.~(\ref{E:acoul}).

Unlike the Coulomb potential in commuting geometry, the effective potential in 
$\theta$ NC is finite at a = 0, where we have

\begin{equation}
\begin{gathered}
  \left\langle 0 \right|\frac{1}
{r}\left| 0 \right\rangle  = \frac{1}
{{2\pi ^2 }}\int {\frac{{d^3 q}}
{{q^2 }}\exp \left[ { - \frac{{\sigma \left( {q^3 } \right)^2 }}
{2} - \frac{{\sigma _p {\mathbf{q}}_ \bot ^2 }}
{2}} \right]}  \\ 
   = \frac{1}
{{2\pi }}\int_{ - 1}^{ + 1} {dc} \frac{1}
{{\sqrt {\sigma _p  - \left( {\sigma _p  - \sigma } \right)c^2 } }} \\ 
\end{gathered} 
\end{equation}

If $\sigma  > \sigma _p $, this becomes

\begin{equation}
\left\langle 0 \right|\frac{1}
{r}\left| 0 \right\rangle  = \frac{1}
{{\pi \sqrt {\sigma  - \sigma _p } }}\sinh ^{ - 1} \left( {\sqrt {\frac{{\sigma  - \sigma _p }}
{{\sigma _p }}} } \right)
\end{equation}

At $\sigma  = \sigma _p $ this becomes $\frac{1}
{{\pi \sqrt {\sigma _p } }}$ , and for $\sigma  < \sigma _p $
we get

\begin{equation}
\left\langle 0 \right|\frac{1}
{r}\left| 0 \right\rangle  = \frac{1}
{{\pi \sqrt {\sigma _p  - \sigma } }}\arcsin \left( {\sqrt {\frac{{\sigma _p  - \sigma }}
{{\sigma _p }}} } \right)
\end{equation}

which is finite at $\sigma  = 0$
and equal to $\frac{1}
{{2\sqrt {\sigma _p } }}$ .

\section{The Charge Density}\label{S:chgden}

The charge density, expressed as

\begin{equation}
\begin{gathered}
  \rho  =  - \frac{1}
{{4\pi }}\nabla ^2 \frac{1}
{r} \\ 
   = \frac{1}
{{4\pi }}\left[ {{\mathbf{p}} \cdot ,\;\left[ {{\mathbf{p}},\;\frac{1}
{r}} \right]} \right] \\ 
\end{gathered} 
\end{equation}

has expectation value in the coordinate states given by

\begin{equation}
\begin{gathered}
  \rho \left( {\mathbf{a}} \right) = \left\langle {\mathbf{a}} \right|\rho \left| {\mathbf{a}} \right\rangle  \\ 
   =  - \frac{1}
{{4\pi }}\nabla ^2 \left\langle {\mathbf{a}} \right|\frac{1}
{r}\left| {\mathbf{a}} \right\rangle  \\ 
\end{gathered} 
\end{equation}

Using Eq.~(\ref{E:potina}) this becomes a Gaussian distribution:

\begin{equation}\label{E:rho}
\rho \left( {\mathbf{a}} \right) = \frac{1}
{{\left( {2\pi } \right)^{\frac{3}
{2}} \sqrt \sigma  \sigma _p }}\exp \left( { - \frac{{\left( {a^3 } \right)^2 }}
{{2\sigma }} - \frac{{{\mathbf{a}}_ \bot ^2 }}
{{2\sigma _p }}} \right)
\end{equation}

\section{The ``Effective''  Electric Field}\label{S:elecfld}

With an effective potential at the point ${\mathbf{a}}$ defined as 

\begin{equation}
\phi \left( {\mathbf{a}} \right) = \left\langle {\mathbf{a}} \right|\frac{1}
{r}\left| {\mathbf{a}} \right\rangle 
\end{equation}

we can define an effective field at ${\mathbf{a}}$ as

\begin{equation}
{\mathbf{E}}\left( {\mathbf{a}} \right) =  - \nabla _a \phi \left( {\mathbf{a}} \right)
\end{equation}

The ``effective'' energy contained in this field may be of some 
interest.  Defining it as

\begin{equation}
U_E  = \frac{1}
{{8\pi }}\int {d^3 a{\mathbf{E}}^2 \left( {\mathbf{a}} \right)} 
\end{equation}

we find

\begin{equation}
\begin{gathered}
  U_E  = \frac{1}
{{8\pi }}\int {d^3 a\left( {\nabla \phi } \right)} ^2  \\ 
   =  - \frac{1}
{{8\pi }}\int {d^3 a\phi \nabla ^2 \phi }  \\ 
   = \frac{1}
{2}\int {d^3 a\phi \rho }  \\ 
\end{gathered} 
\end{equation}

From Eqs.~(\ref{E:potina}) and (\ref{E:rho}) we get, for $\sigma  < \sigma _p $
,

\begin{equation}
U_E  = \frac{1}
{{2\sqrt 2 \pi \sqrt {\sigma _p  - \sigma } }}\arcsin \left( {\sqrt {\frac{{\sigma _p  - \sigma }}
{{\sigma _p }}} } \right)
\end{equation}

At the extreme limit, when $\sigma  = 0$ and $\sigma _p  = \frac{\theta }
{2}$, this becomes $U_E  = \frac{1}
{{4\sqrt \theta  }}$.

\section{The True Electric Field}\label{S:trufld}

Although the integral over all space of the square of the expectation value of the 
electric field operator is finite, this is not true of the operator $
{\mathbf{E}}^2$ itself.  Let $A\left({\mathbf{x}}  \right)$ be any 
operator function of ${\mathbf{ x}}$.  It can be written as 

\begin{equation}
A\left( {\mathbf{x}} \right) = \int {\frac{{d^3 q}}
{{\left( {2\pi } \right)^4 }}\tilde A\left( {\mathbf{q}} \right)\exp 
\left[ {i{\mathbf{q}} \cdot \left( {{\mathbf{\xi}}  + \frac{1}
{2}{\mathbf{\theta}}  \times {\mathbf{p}}} \right)} \right]} 
\end{equation}

where

\begin{equation}
\tilde A\left( {\mathbf{q}} \right) = \int {d^3 a\,A\left( {{\mathbf{x}} + {\mathbf{a}}} \right)\exp \left( { - i{\mathbf{q}} \cdot {\mathbf{a}}} \right)\exp \left[ { - i{\mathbf{q}} \cdot {\mathbf{x}}} \right]} 
\end{equation}

in particular, the operator's space integral can be found from

\begin{equation}
\tilde A\left( 0 \right) = \int {d^3 a\,A\left( {{\mathbf{x}} + {\mathbf{a}}} \right)} 
\end{equation}

The operator $E^2$ can be written as 

\begin{equation}
{\mathbf{E}}^2 \left( {\mathbf{x}} \right) = \frac{1}
{{4\pi ^4 }}\int {\frac{{d^3 q'd^3 q}}
{{q'^2 \left( {{\mathbf{q}} + {\mathbf{q'}}} \right)^2 }}{\mathbf{q'}} \cdot } \left( {{\mathbf{q}} + {\mathbf{q'}}} \right)\exp \left[ { - \frac{i}
{2}{\mathbf{q'}} \cdot \theta  \times {\mathbf{p}}} \right]\exp \left( {i{\mathbf{q}} \cdot {\mathbf{x}}} \right)\exp \left[ { + \frac{i}
{2}{\mathbf{q'}} \cdot \theta  \times {\mathbf{p}}} \right]
\end{equation}

from which it follows that 

\begin{equation}
\int {d^3 a\,} {\mathbf{E}}^2 \left( {{\mathbf{x}} + {\mathbf{a}}} \right) = \frac{1}
{{4\pi ^4 }}\int {\frac{{d^3 q}}
{{q^2 }}} 
\end{equation}

and is infinite.

\section{Conclusions}

In $\theta$ NC geometry the physical operators corresponding to the 
measurement of position, the $x^{i}$ , are non-commuting, but can be 
exoressed in terms of the momentum operators $p^{i}$, and commuting 
``coordinate'' operators $\xi^{i}$. States of minimum dispersion in 
$x^{i}$ can be expressed as Gaussian eigenfunctions of the 
$\xi^{i}$.  These states are the closest one can get in $\theta$ NC 
to a point particle.

The antisymmetric tensor $\theta^{ij}$ defines a vector 
$\theta^{k}$.  The physical distance to a point charge, for fixed 
coordinate distance, depends on the direction relative to 
${\bf{\theta}}$ along which the distance is measured.  The departures 
of the measured Coulomb potential from 1/r are also direction 
dependent.

In commutative geometry, the expectation value of a field operator 
such as the potential, the charge distribution, or the electric field, 
in a Gaussian state located at a point, becomes equal to the value of 
the field at that point when the state's dispersion goes to zero.  
One might therefore consider in $\theta$ NC the expectation of the 
value of the field in a minimum dispersion state centered at some 
point to be the ``effective'' equivalent of the value of the field at 
that point.  We find the effective value of the potential at zero 
position to be finite.  Also the integral over all space of 
the''effective'' electrostatic energy density of a point charge is 
finite, as discussed by Colatto et al.~\cite{colatto} However, the 
``effective'' value involves just one matrix element.  We show that the 
integral over all space of the energy density operator is still 
infinite, just as in commutative geometry.

\section{Acknowledgements}

I would like to thank O. W. Greenberg for many helpful discussions, 
and also the University of Maryland Physics Dept. for their very kind 
hospitality.

\end{document}